\documentclass[useAMS,usenatbib]{mn2e}

\usepackage{bm}
\usepackage{amsmath}
\usepackage{amssymb}
\usepackage{color}
\usepackage{pst-grad}
\usepackage{psfrag}
\usepackage{graphicx}

\newcommand{\mnras}{MNRAS}
\newcommand{\apj}{ApJ}
\newcommand{\apjl}{ApJ}
\newcommand{\nat}{Nature}
\newcommand{\aap}{A\&A}
\newcommand{\araa}{ARA\&A}
\newcommand{\jgr}{J. Geophys. Res.}

\newcommand{\eqb}{\begin{eqnarray}}
\newcommand{\eqe}{\end{eqnarray}}
\newcommand{\diff}{{\rm d}}

\title[Cosmic-ray filamentation instability]{A filamentation instability for streaming cosmic-rays}

\author[Reville \& Bell]{B. Reville\thanks{E-mail:
b.reville1@physics.ox.ac.uk}, A.~R. Bell\\
Clarendon Laboratory, University of Oxford, Parks Road, Oxford OX1 3PU, United Kingdom}

\begin{document}

\date{Accepted 2011 September 26. Received 2011 September 26; 
in original form 2011 July 13}

\pagerange{\pageref{firstpage}--\pageref{lastpage}} \pubyear{2011}

\maketitle

\label{firstpage}

\begin{abstract}

We demonstrate that cosmic rays form filamentary structures in the precursors of supernova remnant shocks 
due to their self-generated magnetic fields. The cosmic-ray filamentation  
results in the growth of a long wavelength instability, and naturally couples the rapid non-linear amplification 
on small scales to larger length scales. Hybrid magnetohydrodynamics--particle
simulations are performed to confirm the effect. The resulting large scale magnetic field may facilitate
the scattering of high energy cosmic rays as required to accelerate protons beyond 
the knee in the cosmic-ray spectrum at supernova remnant shocks. 
Filamentation far upstream of the shock may also assist in the escape of cosmic rays from the accelerator.
\end{abstract}

\begin{keywords}
acceleration of particles -- magnetic fields -- plasmas -- cosmic rays.
\end{keywords}

\section{Introduction}

It is generally accepted that galactic cosmic rays are accelerated in 
supernova remnants. The process of diffusive shock acceleration 
\citep{krymskii77,axfordetal77,bell78a,blandfordostriker78} remains
the most likely mechanism for producing and maintaining the observed spectrum.
There is now a growing wealth of observational evidence supporting this scenario. 
The detection of TeV gamma-ray emission from nearby remnants 
confirms the presence of electrons, and possibly protons, with energies of 
at least $10^{14}$~eV \citep[e.g.][]{hintonhofmann09}. 
In addition, high resolution observations of narrow non-thermal X-ray
filaments at the outer shocks of several 
shell-type supernova remnants favour a model in which the highest energy electrons
are produced directly at the shock, consistent with the predictions of diffusive shock acceleration
\citep[e.g.][]{vinklaming03,bambaetal05,uchiyamaetal07}.
These filaments also provide evidence for strong magnetic field 
amplification in the vicinity of the shock. The generation of strong magnetic turbulence 
is vital for the acceleration of cosmic rays to the knee ($\sim10^{15.5}$~eV) and above
\citep{lagagecesarsky83, belllucek01}. 

While several mechanisms for amplifying magnetic fields to values in excess of the shock compressed
interstellar fields have been proposed, those that result in the transfer of upstream cosmic-ray streaming energy
to the magnetic field are of greatest relevance for diffusive shock acceleration. 
The non-resonant mode first
identified by \cite{bell04} has been demonstrated to grow rapidly, however, the characteristic wavelength
of the amplified field is
predominantly on a length scale shorter than the gyroradius of the driving particles. 
Under certain conditions, other short-wavelength instabilities may dominate
\citep[e.g][]{bret09,riquelmespitkovsky10,lemoinepelletier10,nakaretal11}.
While such instabilities
may be sufficient to explain the large magnetic field values inferred from observations,
in order to facilitate rapid acceleration to higher energies, it is necessary to generate field structure on scales comparable with the gyroradius of the highest energy particles 
\citep{belllucek01,revilleetal08}.

The generation of large scale field structures has been investigated in the
context of filaments, or beams, in \citet{bell05}, where a pre-existing profile for the cosmic ray 
distribution was assumed. In this paper, we demonstrate that the distribution of 
relativistic particles is inherently non-uniform, and that 
filamentation occurs as a natural consequence of cosmic-ray streaming.
A similar phenomenon occurs in laser plasmas whereby photon beams 
filament due to thermal self-focusing in expanding cavities
\cite[e.g.][]{craxtonmccrory84}. We show here that in the case of cosmic-rays,
this process results in the growth of magnetic field on large scales.
The development of the filamentation and large-scale field is investigated analytically
in a two dimensional slab symmetric geometry, and verified
using hybrid particle-MHD simulations. The non-linear feedback between ultra relativistic particles and the background plasma, and the resulting large-scale fields will have important
implications for the acceleration of cosmic rays to energies above the knee in supernova remnants,
and also their escape.

The outline of the paper is as follows. In the next section we develop the analytic model
that describes the cosmic-ray filamentation. It is demonstrated that this introduces a long wavelength
instability in the precursors of supernova remnant shocks. In section \ref{sims_sect},
we report on the numerical code used to investigate the instability, and present simulation results. 
The relevant time and length scales inferred from theory and observations
are addressed in section \ref{compare_sect}.
We conclude with a discussion of the implications for cosmic ray acceleration
and escape of cosmic rays upstream of the shock in supernova remnants in the context of filamentation.

\section{Cosmic ray filamentation}
\label{anal_sect}

Within the diffusion approximation of shock acceleration theory,
a first order anisotropy is introduced in the upstream particle distribution as a result
of the gradient in the isotropic part of the distribution. For a shock front propagating 
in the positive $x$ direction, with velocity $u_{\rm sh}$, the resulting steady state
test-particle solution is \cite[e.g][]{drury83}
\begin{equation}
\label{anisotropy}
 F(x,\bm{p}) = f_0(x,p)\left(1+3\frac{u_{\rm sh}}{c}\cos\theta\right)
\end{equation}
where $f_0(x,p)$ is the isotropic part of the spectrum as measured in the upstream rest-frame,
and $\theta$ is the particle pitch angle with respect to the shock normal.
It is generally understood that the current associated with the anisotropic part of the 
upstream particle distribution drives the growth of MHD instabilities. The resulting 
turbulent magnetic fields mediate the scattering that maintain the cosmic rays' 
quasi-isotropic distribution, ensuring a high probability that particles repeatedly 
cross the shock before escaping.

To investigate the behaviour in higher dimensions, it is convenient to use the Vlasov
equation, which for ultra-relativistic particles can be written in the form
\eqb
\label{vlasov}
\frac{\partial f}{\partial t} + c\frac{\bm{p}}{p}\cdot \bm{\nabla}{f}
+e\left(\bm{E}+{\bm{v}}\times\bm{B}\right)\cdot \frac{\partial f}{\partial \bm{p}}=0
\enspace .
\eqe
In a reference frame in which the upstream cosmic rays are isotropic, 
the distribution function $f(\bm{x},p,t)$ is dependent only
on the length of the momentum vector, not its direction, and it is straightforward
to show that the magnetic field term makes no contribution. It follows that
the only force relevant for calculating the particle distribution
is that due to the local electric field. On average, the upstream cosmic-ray distribution is isotropic in 
the rest frame of the shock \citep{mcclementsetal96}. 
Neglecting the bulk deceleration of the incoming plasma due to the cosmic-ray pressure gradient,
the background plasma in this frame moves towards the shock with velocity
$\bm{u}= -u_{\rm sh}\bm{\hat{x}}+\delta{\bm{u}}$, where $\delta\bm{u}$ are the superimposed
background fluid motions due to the cosmic-ray current, and $\bm{\hat{x}}$ is the unit vector 
along the direction of the shock normal. Conservation of momentum dictates
that these motions are small compared to the 
shock velocity $|\delta{\bm u}|\ll u_{\rm sh}$, and to lowest order
the local electric field is, in the ideal MHD limit,
$\bm{E}=u_{\rm sh}\bm{\hat{x}}\times\bm{B_\bot}$. While this analysis is valid for all shock 
obliquities, we focus here on self-generated magnetic fields $B_\bot$ due to current-driven instabilities.

To investigate the role of filamentation in the plane normal to the direction of the cosmic-ray
streaming, we neglect the cosmic-ray pressure gradient in the precursor, and
restrict the analysis to the case of slab symmetry in the $x$ direction. Introducing the vector potential
$\bm{B}=\bm{\nabla}\times\bm{A}$, the local electric field is
\eqb
\bm{E}={u_{\rm sh}}\bm{\nabla}A_\|(y,z)\enspace,
\eqe
where the scalar potential is $A_\|=\bm{A}\cdot\bm{\hat{x}}$.
Inserting into (\ref{vlasov}), the distribution function, when observed in a reference
frame in which the shock is at rest, evolves according to
\eqb
\frac{\partial f}{\partial t} + c\frac{\bm{p}}{p}\cdot \bm{\nabla}{f}
+e\bm{\nabla}\left(u_{\rm sh}A_\|\right)\cdot \frac{\partial f}{\partial \bm{p}}=0
\eqe
where $u_{\rm sh}A_\|$ plays the role of the effective electric field potential
\cite[e.g.][section 8.17]{kralltrivelpiece}.
On the slowly evolving MHD timescales, the cosmic-ray distribution will progress 
through equilibrium states 
\eqb
\frac{\partial f(\epsilon)}{\partial t }=0\enspace, \enspace
\mbox{ where }\enspace  
\epsilon=p-eu_{\rm sh}A_\|/c\enspace . \nonumber
\eqe
It follows that the phase-space distribution of the cosmic rays
consists of surfaces of equal density on the momentum iso-surfaces
$\epsilon$. Hence, if $\partial f/\partial p <0$, as is almost certainly
the case,
the cosmic-ray number density will be locally larger (smaller) in regions of positive (negative)
$A_\|$. Specifically, if $p\gg eu_{\rm sh}|A_\||/c$, making a Taylor expansion,
the number density as a function of position is 
\eqb
n_{\rm cr}(y,z)=n_0+\frac{eu_{\rm sh}A_\|}{c}\int 8\pi p f_0 \diff p 
\label{crdensity}
\eqe
where we have performed an integration by parts.
Here $f_0$ is the unperturbed part of the spectrum, and $n_0=\int4\pi p^2f_0\diff p$
the associated uniform number density. Note that the correlation with $A_\|$ is dependent
on the choice of orientation. If the upstream instead was chosen to lie in the half plane $x<0$, the 
density and vector potential would anti-correlate. In addition, the correlation is charge dependent,
such that in the precursor, the electrons and protons will anti-correlate. Since the number density
of non-thermal particles is very much less than that of the background plasma, on the 
length scales of interest, charge neutrality is always maintained.

The growth of the magnetic field is driven by the 
resulting cosmic-ray current. Transforming back to the upstream frame, from (\ref{crdensity}),
it follows that the cosmic-ray current is also a function of position
\eqb
\label{jandA}
j_{\rm cr}(y,z)=j_{0}+\chi(A_\|-\langle A_\|\rangle)
\eqe
where the additional term $\langle A_\|\rangle$ has been added to conserve total particle number.
Since lower energy particles are confined closer to the shock, the distribution is expected to fall
off rapidly below a minimum momentum $p_{\rm min}$, where $p_{\rm min}(x)$ 
increases with distance from the shock \citep[e.g.][]{eichler79}. 
Assuming a power-law spectrum $f_0(p>p_{\rm min})\propto p^{-4}$ 
\eqb
\chi=\frac{e^2u_{\rm sh}^2}{c}\int 8\pi pf_0 \diff p =
\frac{e^2n_0u_{\rm sh}^2}{p_{\rm min}c}\enspace .\nonumber
\eqe
It has been implicitly assumed here that the distribution function is gyrotropic such that 
$\bm{j}_{\rm cr}\times\bm{\hat{x}}=0$. This naturally holds on scales larger than the cosmic-ray gyroradius, but also on 
smaller scales provided small-angle scatterings on the background field are sufficiently frequent.
All of these results have been verified using high resolution hybrid simulations (see section \ref{sims_sect}).

The role of a single cosmic-ray filament, or beam, with fixed cosmic-ray current
has been previously investigated in \cite{bell05}.
We have demonstrated here that the cosmic-ray current is in fact filamentary in general. This will alter the 
growth of plasma instabilities, and in the next section we investigate this effect.
 
\begin{figure*}
\begin{center}
 \hbox{
\hspace{0.1\textwidth}
 \includegraphics[width=0.35\textwidth]{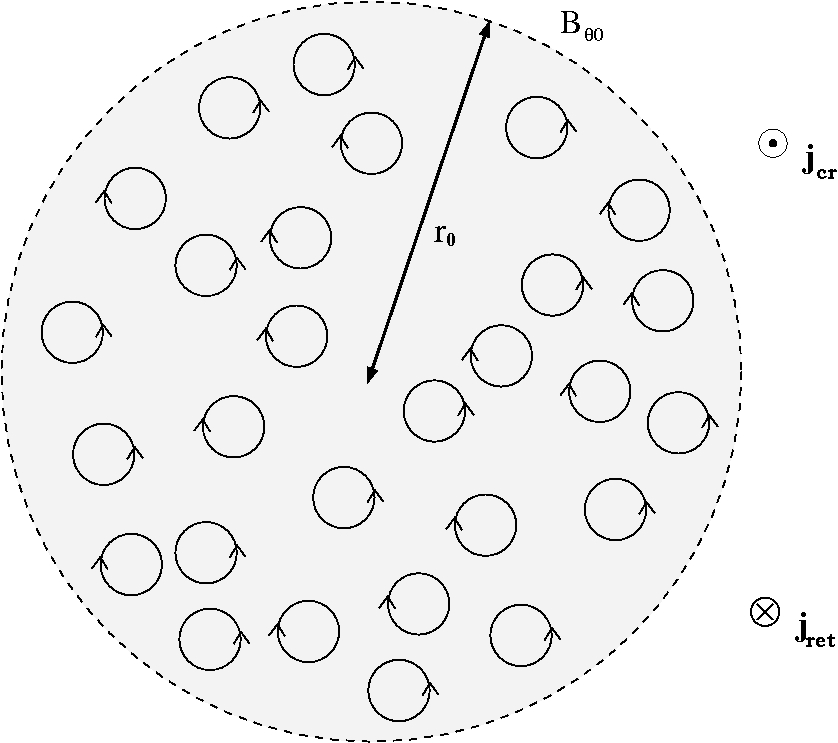}
\hspace{0.1\textwidth}
 \includegraphics[width=0.35\textwidth]{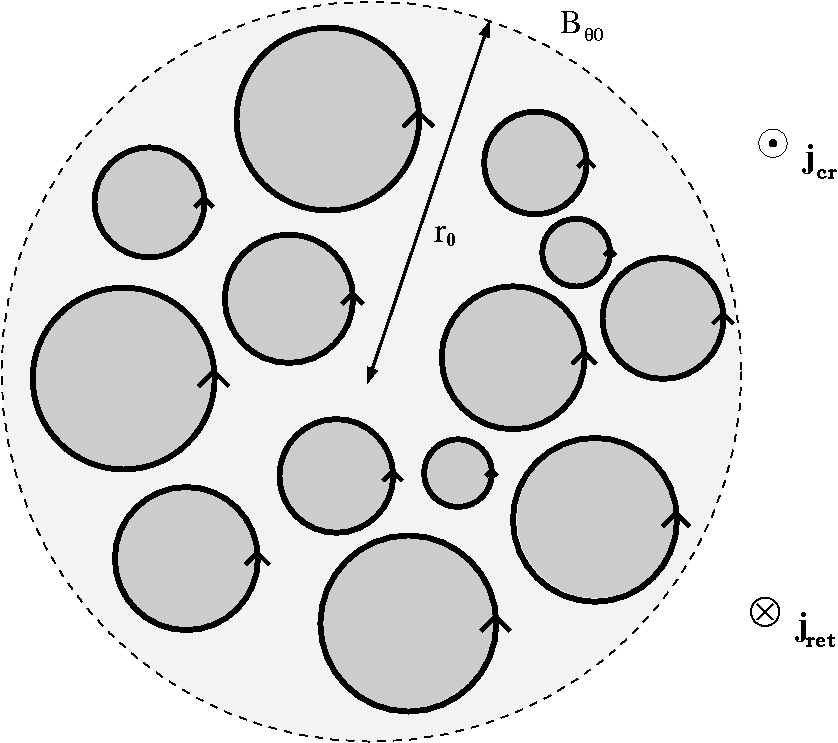}
\rput(-9,0){Early Times}
\rput(-1,0){Late Times}
}
\end{center}
\caption{Illustration of the behaviour on different length scales. $\bm{j}_{\rm cr}$ is the cosmic ray 
current, and $\bm{j}_{\rm ret}=(\nabla\times\bm{B})/\mu_0-\bm{j}_{\rm cr}$ the return current carried by 
the background plasma. The small scale circles represent 
the expanding loops of magnetic field of a particular handedness. At early times (left), the magnetic field 
contains loops of both orientation having comparable strength, and the cosmic ray current is approximately uniform.
Since only loops with favourable orientation can grow, in this example counter-clockwise, the cosmic rays
are focused into these expanding loops, while clockwise loops contract. At later times (right)
the large scale magnetic field $B_{\theta0}$ on scale $r_0$ enclosing the smaller expanding loops,
will have increased.
}
\label{fig1}
\end{figure*}

\subsection{Filament growth}

The amplification of magnetic fields in the precursors of supernova remnant shocks is 
a multi-scale problem. 
While the non-linear growth of magnetic field, driven by cosmic-ray
currents has been demonstrated via numerical simulations
\citep[e.g.][]{bell04,ohiraetal09,riquelmespitkovsky09,stromanetal09}, 
most simulations have focused exclusively on field amplification
on small scales. This is primarily
due to the fact that, on larger length scales, it becomes necessary to include the dynamics of
the cosmic rays \citep{lucekbell00}. We use the analysis of the previous section to self-consistently model 
the interaction between the background plasma and the cosmic rays.

Following \cite{bell05}, we analyse the MHD equations with an external cosmic-ray current, 
neglecting the role of pressure gradients and magnetic tension
\eqb
\label{MHDB}
\frac{\partial \bm{B}}{\partial t}&=&\bm{\nabla}\times(\bm{u}\times\bm{B})\\
\label{MHDu}
\rho\frac{\diff \bm{u}}{\diff t} &=& -\bm{j}_{\rm cr}\times\bm{B}
\eqe
valid in the long wavelength approximation.
Also, from (\ref{MHDB}), the vector potential satisfies
\eqb
\label{dAdt}
\frac{\partial \bm{A}}{\partial t} = \bm{u}\times(\bm{\nabla}\times\bm{A}) .
\eqe

Returning to the 2D analysis of the previous section, i.e. zero gradient in the direction of
cosmic-ray streaming, it follows that the parallel component
of the vector potential is constant for a particular fluid element
\eqb
\label{Aconst}
\frac{\diff A_\|}{\diff t} = 0\enspace .
\eqe
This has a number of important consequences. From (\ref{jandA}), the cosmic ray
current will increase in regions of large $A_\|$. Considering an 
idealised axisymmetric system, with maximum $A_\|$ is centred on the origin,
$B_\theta=-\partial A_\|/\partial r>0$, at least locally. Hence, the resulting 
$-\bm{j}_{\rm cr}\times\bm{B}$ force acts to push the plasma radially outwards. This spreads the
region of large $A_\|$, thus focusing more cosmic rays into the filament, leading to a 
run-away instability. In a 2D slab symmetric geometry, this results
in the spreading out of flat table-top structures with large $A_\|$ surrounded by regions of 
negative $A_\|$, with large gradients in between. This is similar to the picture
presented in \citet{bell05} section 3, where the cosmic ray current was fixed,
and the growth rate for the expansion of cavities was found to be
\eqb
\label{nrgrowth}
\Gamma_{\rm nr}=\left(\frac{j_{\rm cr}B_{\theta}}{r\rho_0}\right)^{1/2}\enspace ,
\eqe
with $r$ the radius of the cavity, and $B_{\theta}$ the magnetic field 
strength on that scale.
Here, the focussing of the cosmic rays into the cavities will enhance the growth 
rate as compared with the constant current case, since $j_{\rm cr}$ is larger in the filaments.

For growth on small scales the orientation of the magnetic field must be favourable.
Considering a field configuration such that at early times, an equal number of small-scale loops of
both polarisations are randomly located within a circle of radius $r_0$, as 
shown in Figure \ref{fig1}, the effect of the cosmic ray current is to expand loops 
of one orientation and contract the other. The net result is that
the small scale loops are predominantly of a single polarisation at late times. This 
corresponds to a net current in the direction of the cosmic-ray streaming when averaged 
over the area enclosed by $r_0$, i.e. $\langle\nabla\times\bm{B}\cdot\bm{\hat{x}}\rangle\neq0$. 
As the total current enclosed by $r_0$ increases, 
the magnetic field $B_{\theta0}$ on this scale must likewise increase. Unlike the small scale 
fields, however, the growth will be independent of orientation.

To quantify the above simple picture, we combine (\ref{jandA}), (\ref{MHDB}) 
and (\ref{MHDu}), together with the equation
${\partial \bm{A}}/{\partial t}=\bm{u}\times\bm{B}$, to give
the following expression for the evolution of the filamentation 
\eqb
\label{MHDj}
\frac{\partial^2 j_{\rm cr}}{\partial t^2}=\frac{\chi B_\bot^2}{\rho}j_{\rm cr}+
\left(\left[ (\bm{u}\cdot\bm{\nabla})\bm{u}\right]\cdot\bm{\nabla}\right)j_{\rm cr}
-(\bm{u}\cdot\bm{\nabla})\frac{\partial j_{\rm cr}}{\partial t} \enspace.
\eqe

The second two terms on the right hand side of equation (\ref{MHDj}) 
represent the advection of $A_\|$ with the flow. On small scales, where the velocity 
gradients are steep, these terms dominate over the first term. However, since the continued
expansion of small scale loops is eventually inhibited by neighbouring cavities, 
\citep{bell04,revilleetal08}, on sufficiently large length scales the first term will dominate.
The ordering of these terms will be verified in section \ref{compare_sect}.

Neglecting the last two terms in equation (\ref{MHDj}), we find the following
growth rate for the filamentation instability
\eqb
\label{filgrowth}
\Gamma_{\rm fil}=\sqrt{\frac{\chi B_\bot^2}{\rho}}=\eta
\left(\frac{u_{\rm sh}}{c}\right)^2\left(\frac{U_{\rm cr}}{\rho u_{\rm sh}^2}\right)^{1/2}
\frac{eB_\bot^{\rm rms}}{\gamma_{\rm min} m}
\eqe 
where $U_{\rm cr}$ is the cosmic-ray energy density, $\gamma_{\rm min}=p_{\rm min}/mc$
the Lorentz factor of the lowest energy cosmic rays driving the instability
(i.e. those satisfying $p_{\rm min}c\gg eu_{\rm sh}A_\|$), and $\eta$
is a numerical factor that depends on the shape of the cosmic-ray spectrum. 
For a spectrum $f\propto p^{-4}$ in the momentum interval $(p_{\rm max}>p>p_{\rm min})$, 
this parameter is $\eta=1/\sqrt{\ln(p_{\rm max}/p_{\rm min})}$.
The growth rate is scale independent, and depends only on the root mean square of
the perpendicular magnetic field 
enclosed on that scale, as expected from the qualitative description above.
Since the non-resonant mode discussed in \citet{bell04} has a growth rate that decreases 
monotonically with increasing wavelength, the filamentation must dominate the amplification 
of magnetic field on some scale. However, for the growth rate of the filamentation instability
to be sufficiently rapid to influence the scattering of high-energy cosmic rays, 
the mean squared magnetic field on small scales must be amplified 
to values well in excess of the ambient field. 
Thus, the filamentation instability can be considered as a bootstrap to
the non resonant instability described in \citet{bell04} and \citet{bell05}. 
The necessary conditions for the filamentation to play an important role 
are discussed in detail in section \ref{compare_sect}.

The transfer of magnetic energy from small scales to longer wavelengths in the context of diffusive shock
acceleration has previously been suggested to occur via an inverse cascade 
\citep[e.g.][]{pelletieretal06,diamondmalkov07}. Whether this cascade can bridge
the large separation of scales remains uncertain. The mechanism described here presents a different 
approach where the coupling of the scales is mediated by the filamentation. The coupling
of small and large scale magnetic fields has also been found in simulations of  
sheared flows with small scale turbulence \citep{yousefetal08} in the context of mean-field 
dynamo theory. This approach has also recently been applied to case of precursors
with an external cosmic-ray current \citep{bykovetal11,schurebell11}.

\section{Numerical simulations}
\label{sims_sect}

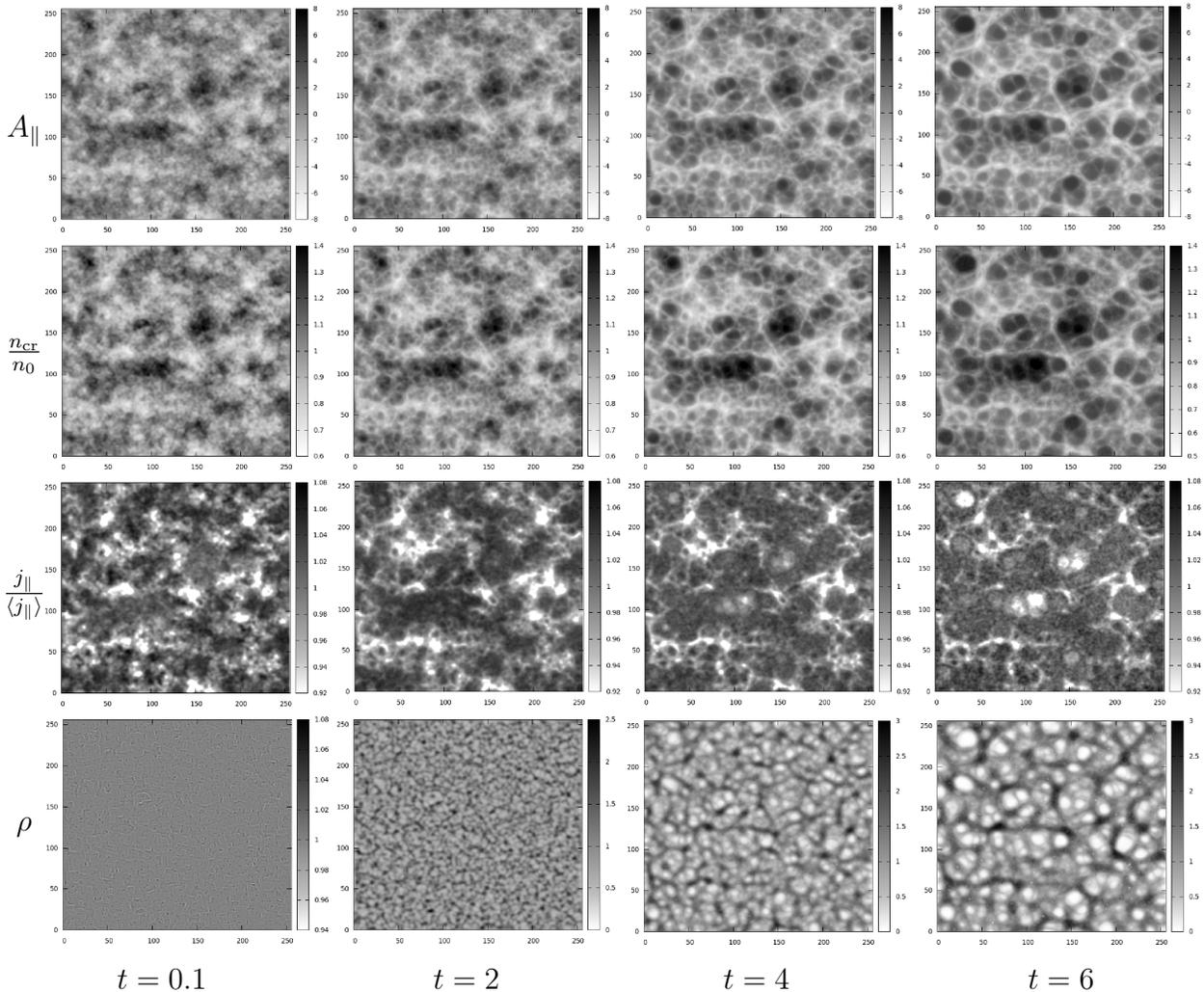
\begin{figure*}
~~\includegraphics[width=0.9\textwidth]{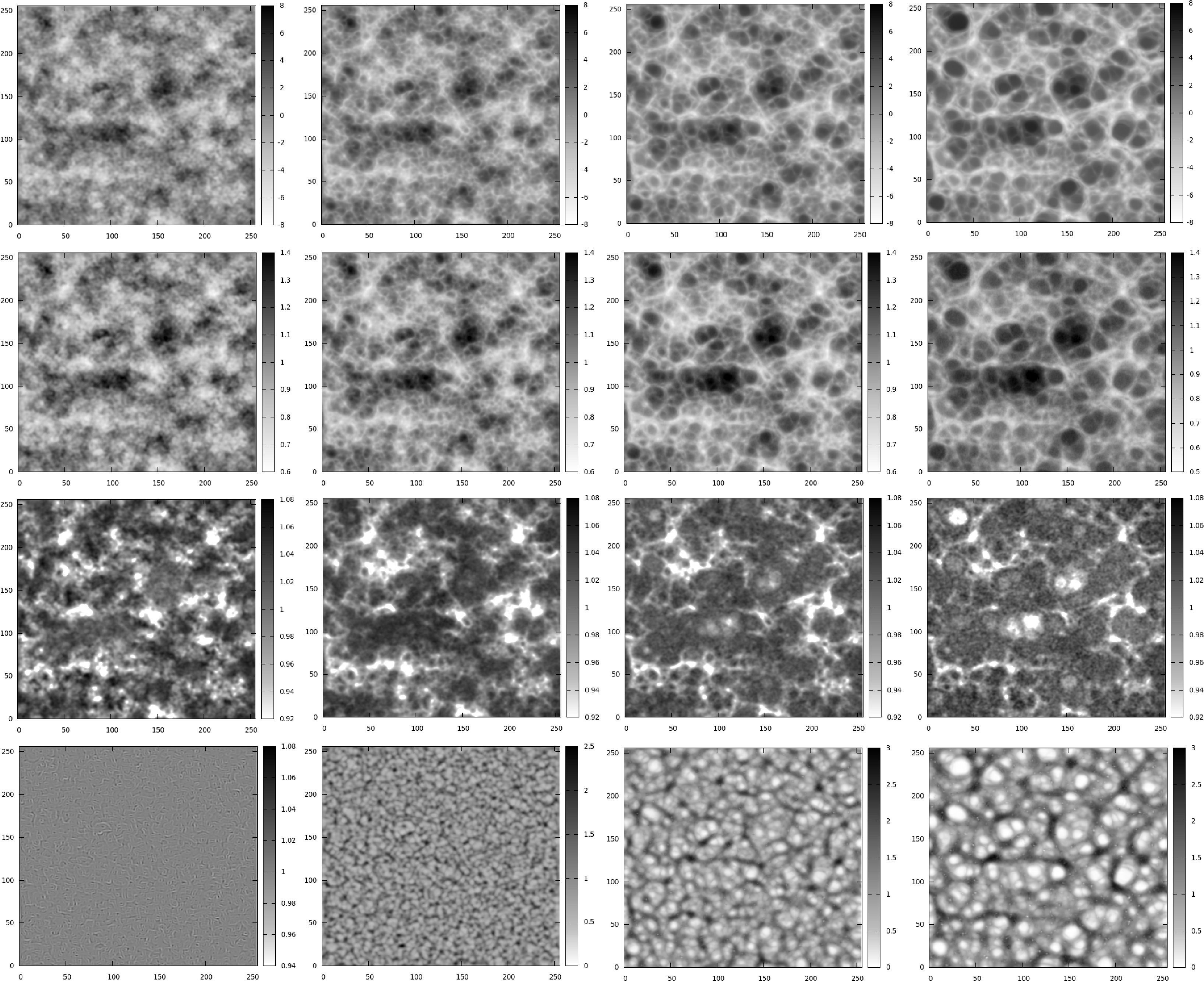}
\vspace{0.8cm}
\caption{Time dependent growth of the filamentation.
$y{\rm -}z$ plots of $A_\|$, $n/n_0$, $j_\|/\langle j_\|\rangle$ and $\rho$ at $t=1,2,3,4$. Cosmic
rays current is in the positive $x$ direction (out of the plane). Details of the simulation 
parameters are found in the text.}
\label{fig2}
\rput(-8.2,13.5){{\Large $A_\|$}}
\rput(-8.2,10.4){{\Large $\frac{n_{\rm cr}}{n_0}$}}
\rput(-8.2,7.1){{\Large $\frac{j_\|}{\langle j_\|\rangle}$}}
\rput(-8.2,3.9){{\Large $\rho$}}
\rput(-6.3,1.8){{\Large $t=0.1$}}
\rput(-2.1,1.8){{\Large $t=2$}}
\rput(1.9,1.8){{\Large $t=4$}}
\rput(6.1,1.8){{\Large $t=6$}}
\end{figure*}

Numerical simulations are performed to verify the analysis of the previous section.
To investigate these processes, it is necessary to have a kinetic description of the cosmic-rays. 
A code has been developed similar to that described by \citet{zacharycohen86} and \citet{lucekbell00}, 
where the background plasma is treated as an MHD fluid and the 
cosmic rays are treated using a particle-in-cell (PIC) approach.
This method is appropriate for modelling plasmas in which there 
exists a large separation between the relevant length and time scales associated with the thermal and 
non-thermal particles' kinetics. Unlike full PIC simulations, which solve Maxwell's equations directly,
our simulations use the magnetic and electric fields determined from the MHD equations, i.e..
the electric field is determined from ideal MHD $\bm{E}=-\bm{u}\times\bm{B}$, and the displacement current
is neglected.
The cosmic-rays are evolved in these fields by integrating the relativistic equations of motion
using the Boris method 
\citep[e.g.][]{birdsalllangdon} and the resulting current and charge densities are
deposited on the grid at each particle update using the approach of \citet{umedaetal03}.
Since the particles typically evolve on a shorter time-scale than the background MHD, it is necessary to 
integrate particle trajectories over several sub-cycles within each MHD update 
\citep{zacharycohen86}. While this  
multiplies the effective number of particles in the simulation, it can also smooth out 
phase correlations associated with the cosmic-ray current \citep{zacharyetal89,lucekbell00}.
The coupled MHD -- cosmic-ray equations are solved using a finite-volume Godunov scheme, 
as described in \citet{revilleetal08}, with the self-consistent inclusion of temporal and spatially 
varying current and charge densities $j_{\rm cr}$, $Q_{\rm cr}$.

The simulations are run using a 2D slab symmetric geometry, as in the analysis of
the previous section, with a periodic grid in $y-z$ plane, and the cosmic
ray anisotropy directed out of the simulation plane in the positive $x$-direction. 
The particles are initialised as a mono-energetic distribution with a net drift velocity.
Maintaining gyrotropy in the simulations and thus minimising $j_\bot$ due to particle noise, 
requires that a large number of particles per cell be used. For the results shown in this
paper, the particles are initialised with $1024$ per cell. 
To minimise the noise in the particle
distribution at $t=0$, the discrete particle momentum vector in spherical coordinates, 
$\phi_i$ and $\mu_i$, are chosen in an ordered manner, such that they satisfy the required 
distribution globally to a high degree of accuracy. Here $\mu=+1$ corresponds to the positive $x$
direction. This is achieved by choosing the azimuthal
angles with uniform spacing as $\phi_i= 2 \pi M(-1)^ii/(N-1)$ for $i=0..N$, where $N$ is the total 
number of particles, and $M$ is the number of complete rotations through $2\pi$ required. We found best results 
for $M=3$. Having too many or too few revolutions in $\phi$ at pitch angles $\mu$ close to 
zero can result in a numerically introduced anisotropy in the perpendicular direction
after only a relatively small number of time steps. The particle cosines $\mu_i$ are
chosen between $\mu_i=-1..1$ with decreasing spacing $\Delta\mu_i$ 
such that the total distribution has a net drift with the required velocity.
Finally the particles are scattered
in the $y-z$ plane using mixed radix bit-reversed fractions \citep{birdsalllangdon}. 

The two-dimensional magnetic field is initialised by taking a sum of Fourier modes in $A_\|$
sampled from a user-defined spectrum, having random wave vectors $\bm{k}$ in the $y-z$ plane
\citep[see e.g.][]{giacalonejokipii99}.
For the results shown in Figure \ref{fig2}, we used a 2D Fourier spectrum 
$A_\|^2(k)\propto1/[1+(kL_{\rm c})^{3}]$, such that perpendicular magnetic field 
will have a power spectrum peaking close to the length $L_{\rm c}$. 
Several different forms for the power spectrum have been used, and 
the general results are the same. The computational grid is
a $512\times512$ square mesh with periodic boundary conditions in the $y$ and $z$
directions. The grid resolution is $\Delta x = 0.5$, 
where dimensionless length units are chosen such that $mc/eB_0=1$ with 
$B_0$ the mean magnetic field strength out of the plane. The Fourier modes were selected with 
uniform logarithmic spacing on the interval $4<k/2\pi<256$ with $L_{\rm c}=16$, such that most of the
magnetic field structure is on scales much smaller than the size of the box.
The total magnetic field is then calculated by taking the curl, and projecting onto
the grid using central differencing $\bm{B} = B_0\bm{\hat{x}}+\bm{\nabla}\times A_\|\bm{\hat{x}}$. 
This guarantees that the field is divergence free.

The thermal background is initialised at rest with uniform density and pressure. The 
cosmic-ray current 
and charge densities are also initially uniform, to within noise levels on the grid. 
A rather large shock velocity of $u_{\rm sh}=0.5 c$ is used to enhance the anisotropy. 
If the departure from isotropy is too small, a much larger number of particles are 
required to accurately model the perturbations to the cosmic-ray current. Since the
shock velocity only appears in the determination of the cosmic ray anisotropy, the
use of the non-relativistic MHD equations remains valid.  
The numerical parameters are chosen to represent
values that might be expected in a young supernova remnant: $mn_{\rm cr}/\rho_0=10^{-6}$,
$v_{\rm A}/c=5\times10^{-4}$. The 
particle momentum is chosen such that the gyroradius is resolved in the simulation box,
and for the results shown in Figure~\ref{fig2}, the particle gyroradius
is $r_g=20(B/B_0)^{-1}$. We note that with this low minimum cosmic ray energy and high shock 
velocity, the total cosmic ray energy
density is $U_{\rm cr}/\rho_0 u_{\rm sh}^2\approx10^{-4}$, which is considerably smaller than
what is expected in young supernova remnants. These parameters 
have been chosen such that the various physical processes can be easily identified in the 
simulations, although could also indicate that the effect should exist even in shocks that are
accelerating very inefficiently. In reality, for young supernova remnants, the shock velocity would 
be an order of magnitude smaller and the typical cosmic ray energy several orders of magnitude larger,
where $U_{\rm cr}/\rho_0 u_{\rm sh}^2$ can be as large as $30\%$ \citep[e.g.][]{malkovdrury01}. 
Simulations with a much higher energy density resulted in 
extremely rapid evacuation of zero density cavities in the background fluid
which provided insufficient time to study the growth of magnetic field.

\subsection{Results}

\begin{figure}
 \includegraphics[width=0.45\textwidth]{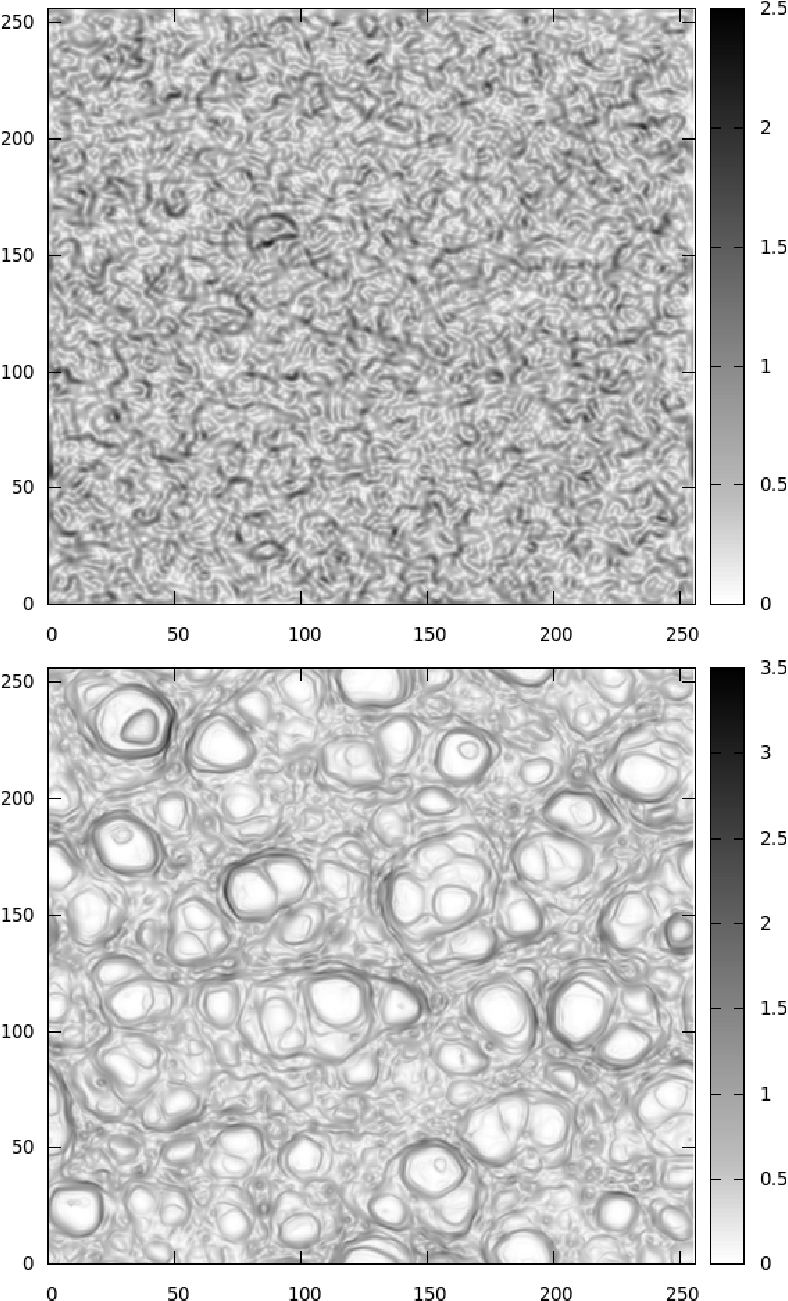}
\caption{Plot of $|B_{\bot}|$ at beginning ($t=0$) and end ($t=8$) of the simulation.
Regions of strong magnetic field wrap around the cosmic ray filaments where the gradients
in $A_\|$ are steep. Cavities are found to develop on a range of scales, in some cases with 
larger loops enclosing smaller loops.
}
\label{fig3}
\end{figure}

The evolution of $A_\|$, $n_{\rm cr}$, $j_\|$ and $\rho$ are shown in Figure~\ref{fig2}.  
The correlation between the vector potential and cosmic ray number density, 
as represented by $n_{\rm cr}$, can clearly be seen. 
The cosmic ray filamentation also results in the creation and expansion of low density cavities in the 
thermal plasma. There also exists a correlation between the cosmic-ray current $j_\|$ and the potential
$A_\|$, although the features are not as sharp as with $n_{\rm cr}$. 
As shown in the previous section, the correlation between $A_\|$ and 
$n_{\rm cr}$ is due to fluctuations in the isotropic component $f_0$. 
The current, on the other hand, depends on higher order expansions in the 
anisotropy of the cosmic ray distribution, and is much more difficult to 
capture numerically, but is clearly evident at early times. As the simulation 
progresses, the net streaming of cosmic rays is gradually reduced. This is always 
to be expected in simulations of this type \citep[e.g.][]{lucekbell00}, 
since the work done on the background plasma by the cosmic rays is $\delta W=-\bm{j}_{\rm cr}\cdot\bm{E}$, 
i.e. in order to extract energy from the cosmic rays, the background plasma must generate 
an opposing electric field. 
Thus the bulk cosmic-ray drift motion will be gradually decelerated,  
causing the correlation between $j_\|$ and $A_\|$ to become weaker as the simulation progresses.
The opposing electric field will be strongest in the walls surrounding the cavities, which can account 
for the anti-correlations found on small scales around regions of large $A_\|$ at late times.
The correlation on large scales is nevertheless evident even at the end of the simulation,
despite a 13\% reduction in the net streaming velocity. In a real system, the cosmic-ray streaming
is continually fed by the cosmic-ray pressure gradient, an effect that can not be captured with our
numerical approach.

\begin{figure}
 \includegraphics[width=0.45\textwidth]{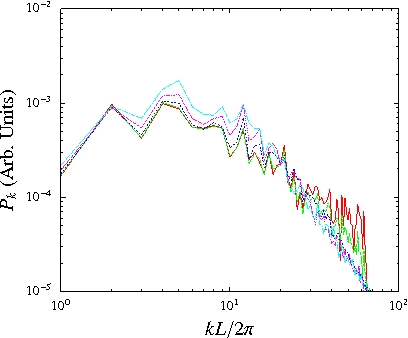}
\caption{Fourier power spectrum of $B_\bot$ at $t=0,2,4,6,8$, in ascending order for
$kL/2\pi<10$, where $L$ is the length of the sides of the simulation grid. 
}
\label{fig4}
\end{figure}

The growth of tabletop features in the 2D plots of $A_\|$ is also observed.
As discussed in section \ref{anal_sect},
in two dimensions, the value of $A_\|$ cannot increase, but regions where $A_\|$ is large 
spread out to form the features seen in the top row of Figure \ref{fig2}. 
It follows from (\ref{jandA}) that the same must hold for the cosmic ray current. 
In a three dimensional situation, any gradients in the $x$ direction will act as a 
source term in (\ref{dAdt}), allowing the possibility of amplifying $A_\|$, resulting in continued 
self-focusing of the cosmic rays. This may have important implications for the upstream escape of 
cosmic rays from supernova remnants. We discuss this further in section \ref{disc_sect}.

The magnetic field at the beginning and end of the simulation are shown in Figure \ref{fig3}.
As found in previous simulations, the regions of strongest magnetic field are concentrated 
in walls surrounding the low density cavities \citep{bell04,bell05,revilleetal08}.
The cavities are produced by the $-\bm{j}_{\rm cr}\times\bm{B}_\bot$ force expanding low density cavities
with the magnetic field frozen in to the background plasma. The largest loops appear to be on 
scales of $r\sim L/10$ where $L$ is the size of the simulation box. To investigate the growth of 
magnetic field on larger scales than this, we perform a Fourier analysis of the field. 
The evolution of the power spectrum for $B_\bot$ is shown in Figure \ref{fig4}. It can be seen that
structures on all wavelengths $kL/2\pi<20$ grow monotonically with time. Since a lot of this structure
is due to the expanding cavities, we focus on the growth of modes with $kL/2\pi<10$, which appear 
to grow almost scale independently, as expected for the filamentation instability. The average measured 
growth rate is $\sim0.25\Gamma_{\rm fil}$, assuming $u_{\rm sh}=0.5 c$ in equation (\ref{filgrowth}).
As already mentioned, in the simulations, the cosmic-ray drift velocity is reduced as the simulation 
progresses which may account for this reduction in comparison with the theoretically predicted growth rate,
which assumed $u_{\rm sh}$ fixed. The resulting effect on the $j_\|$ -- $A_\|$ correlation
may also contribute. 
 
The simulations were terminated when the plasma density fell below a certain threshold. In the 3D simulations of \cite{bell04}, the expanding cavities can merge as field lines slide over one
another, a feature that cannot be reproduced in 2D. Future 3D simulations will allow longer time evolution, and address the issue of saturation of the large scale magnetic field structure.
 
Finally, we emphasise that the primary difficulty in the numerical identification of the instability, 
is choosing parameters such that the filamentation instability dominates on sufficiently small 
length scales, in comparison with the size of the simulation box.
Ultimately, due to the memory limitations, we are left with a
relatively small dynamical range for which a Fourier analysis can be 
carried out, approximately one order of magnitude.
In future simulations, using a different numerical technique, we hope to extend the dynamical range by
at least an order of magnitude, and furthermore, to investigate the role of filamentation in three dimensions.

\section{Application to diffusive shock acceleration}
\label{compare_sect}

Observations of synchrotron X-ray emission in the vicinity of the shock suggest the presence of magnetic fields 
on the order of $100\mu$G or even larger in several supernova remnants. Recent observations have also identified
the presence of a precursor in SN1006, where it is suggested that the 
implied field strengths are larger than the typical interstellar value
\citep{rakowskietal11}, providing tentative evidence for field amplification due to the presence of cosmic rays. 
If these large fields are generated via cosmic-ray
streaming, the mechanism must account for amplification from typical seed fields $B_{\rm ISM}\sim3-5\mu$G.
Numerical simulations of cosmic-ray current driven instabilities on small scales 
suggest non-linear amplification of the fields to values
$B_\bot^{\rm rms}\sim30 B_{\rm ISM}\approx 100\mu$G \citep{bell04,riquelmespitkovsky09}. 
This amplification is believed to occur far upstream
where only the highest energy particles are interacting with the interstellar medium turbulence
\citep{zirakashviliptuskin08,revilleetal09}. The characteristic length scale of the 
fields produced by the non-resonant instability is on too small a scale 
to reduce the mean free path 
$\lambda_{\rm MFP}$ of the highest energy cosmic rays, as required to 
accelerate them beyond the Lagage-Cesarsky limit \citep{lagagecesarsky83}. To increase the acceleration 
rate of the highest energy cosmic rays, the diffusion coefficient $\kappa\approx\lambda_{\rm MFP}c$
must be significantly reduced below its Bohm limit in the pre-amplified field, where the mean free path is equal to
the gyroradius $\lambda_{\rm MFP}=r_{\rm g}$. If the small scale fields are indeed amplified to the 
levels inferred from observations far upstream of the shock via the non-resonant instability, 
as we show here, the filamentation instability can grow on a sufficiently short time scale to
have an appreciable influence on the diffusion of the highest energy cosmic rays.

From the expression for the growth rate given in equation (\ref{filgrowth}), it can be seen that 
the filamentation instability operates most effectively in strongly amplified small scale turbulence,
and when the cosmic-rays driving the instability have lower minimum energy. 
However, if the energy of the cosmic-rays driving the filamentation instability is too 
small, the particles will be trapped. This condition, as derived in section \ref{anal_sect}
is $pc\gg eA_\|u_{\rm sh}$. Using the fiducial values from \citet{bell04} (Equation (21)) 
$k_{\rm max}^{-1}\approx2\times 10^{13} {\rm m}$, the filamentation operates provided 
\eqb
\label{Econd}
&&E_{\rm min} \gg eA_\|u_{\rm sh} \sim e k_{\rm max}^{-1} B_\bot u_{\rm sh}
 \nonumber \\
&&\approx
10^{12} \left(\frac{B_\bot}{100 \mu{\rm G}}\right)\left(
\frac{k_{\rm max}}{2\times10^{13}{\rm m}}\right)^{-1}
\left(\frac{u_{\rm sh}}{10^7 {\rm m/s}}\right) ~{\rm eV}. 
\eqe
TeV $\gamma$-ray observations of most historical supernova remnants provide 
conclusive evidence for the presence of cosmic rays, either protons or electrons, that satisfy this
condition. 

To investigate which mechanism determines the transport properties of the highest energy
cosmic rays in the precursor, we compare the growth rate of the filamentation instability
to that of the streaming instability given in \citet{bell04},
which has a growth rate,
\eqb
\Gamma_{\rm nr}=\sqrt{\frac{j_{\rm cr}B_0k}{\rho}}  \enspace \mbox{   for    } \enspace 
k_{\rm max}>k>r_{\rm g}^{-1}\enspace .
\eqe
This expression is equivalent to equation (\ref{nrgrowth}) on
replacing $B_{\theta}/r$ by $kB_0$ \citep[cf.][]{bell05}. For $k<r_{\rm g}^{-1}$ ion-cyclotron resonance 
takes over and the growth rate steepens $\propto k$. Hence the growth rate $\Gamma_{\rm nr}$ can be considered 
an upper limit for all $k$.

Since the magnetic field amplified by the non-resonant instability is on too small a scale to 
effectively scatter the highest energy cosmic rays driving the growth, i.e. those with $E\lesssim E_{\rm max}$, 
these particles will continue to gyrate about the mean field $B_0$. On the scale of the gyroradius
of these particles $k=eB_0c/E_{\rm max}$, the ratio of the growth
rate of the filamentation instability, equation (\ref{filgrowth}), to the non-resonant instability is 
\eqb
\frac{\Gamma_{\rm fil}}{\Gamma_{\rm nr}}\approx \frac{B_\bot^{\rm rms}}{B_0}
\sqrt{\frac{u_{\rm sh}}{c}\frac{E_{\rm max}}{E_{\rm min}}}\enspace ,
\eqe
where $E_{\rm min}$ is the corresponding energy dominating the cosmic-ray current.  
We note here that the small scale fields are amplified over a distance 
much less than the scale-height of the precursor at the outer extremity of the precursor 
$\approx \kappa(E_{\rm max})/u_{\rm sh}$, suggesting that ${E_{\rm max}}/{E_{\rm min}}$
should not greatly exceed unity.
Thus, provided the small scale fields can be driven to non-linear values, 
the filamentation instability will play the dominant role in generating the fields required to
scatter the highest energy cosmic rays in supernova remnants. 

The two growth rates are equal when
\eqb
\label{rcond}
kr_{\rm g}=\frac{u_{\rm sh}}{c}\frac{\langle B_\bot^{2}\rangle}{B_0^2}
\eqe
where $r_g=E_{\rm min}/eB_0c$. This indicates, to order of magnitude, 
the length scale above which the filamentation dominates over the non-resonant mode. 
Substitution of the parameters used in the simulations suggests the transition occurs at
$kL/2\pi\sim 10$, in agreement with what was found. In addition, comparing the terms in 
equation (\ref{MHDj}), it is readily seen that (\ref{rcond}) corresponds to the scale on 
which the first term becomes comparable with the other terms. 

To demonstrate the important role played by the filamentation instability, we calculate the growth rate
using typical parameters for young supernova remnants.
Assuming the magnetic field is amplified initially on small length scales 
to a level comparable with those inferred from observations, the typical time-scale 
for growth of magnetic field on long wavelengths by the filamentation instability can be as short as 
\eqb 
\Gamma_{\rm fil}^{-1}\approx 50\left(\frac{\eta}{0.2}\right)^{-1}
\left(\frac{U_{\rm cr}/\rho u_{\rm sh}^2}{0.1}\right)^{-1/2}\left(\frac{u_{\rm sh}}{10^7{\mbox{m/s}}}\right)^{-2}
&&\nonumber\\
\times
\left(\frac{B_\bot^{\rm rms}}{100\mu{\rm G}}\right)^{-1}\left(\frac{E_{\rm min}}{10^{14}{\rm eV}}\right) &{\rm yrs}&
\eqe  
The value of $E_{\rm min}$ driving the growth is the largest uncertainty. At the onset of the filamentation,
provided the lower energy cosmic rays satisfy the condition (\ref{Econd}), the growth can be extremely 
rapid. As the magnetic fields evolve, the scale of the filaments becomes comparable to the gyroradius
of the lower energy particles, such that $E_{\rm min}$ should remain large. 
Saturation may occur when the high energy 
particles become trapped on the self-generated large scale fields. This will almost certainly 
affect the diffusion of cosmic rays and may even alter the 
transport properties of particles at different energies,
which can influence the shape of the 
spectrum \citep{kirketal96}. 
Future simulations in three dimensions will help to elucidate this process further.

\section{Discussion}
\label{disc_sect}

While the filamentation of photon or high energy electron beams in laboratory laser plasma
experiments is a well studied phenomenon \cite[e.g.][]{craxtonmccrory84}, its analogy with cosmic rays has been largely overlooked.
In this paper, it has been demonstrated both analytically and confirmed with numerical simulations,
that the filamentation of cosmic rays is an important process that can occur in the precursors of 
supernova remnants shocks where diffusive shock acceleration is taking place.

In addition, we have identified a mechanism for amplifying magnetic field on large length scales
as a result of the filamentation. The process provides a natural mechanism to couple the rapid growth of 
magnetic field on small scales, as driven by the non-resonant instability \citep{bell04}, to length scales 
comparable to, or larger than, the gyroradius of the particles driving this instability, avoiding the need for an inverse-cascade. The growth-time 
for this instability can operate on time scales as short as a few years, provided the small scale 
fields are amplified to a sufficient level. The reason for the short growth time as compared with previous 
calculations of linear dispersion relations, is that the instability 
develops in non-linear small scale magnetic fields with $\delta B^2/B_0^2 \gg 1$, unlike streaming instabilities 
that typically consider the growth of weak Alfv\'enic perturbations in the interstellar medium, 
where $\delta B^2/B_0^2 \ll 1$. 

This has immediate implications for the maximum energy to which cosmic rays can be accelerated at
supernova remnant shock fronts. The amplification of magnetic turbulence on all scales, significantly beyond the limits
of quasilinear theory, remains the most likely possibility for accelerating cosmic rays above the
knee \citep[e.g.][]{belllucek01,kirkdendy01}. Using MHD simulations with a constant external cosmic-ray current, 
\citet{revilleetal08} demonstrated that the non-resonant self generated turbulence, 
reduced the diffusion coefficient of test particles 
significantly below the corresponding Bohm value in the pre-amplified field. However, 
due to the limited dynamic range of these simulations, the diffusion coefficient for particles 
with gyroradii larger than the typical structures in the field, 
converged to the small-angle scattering limit, where the mean free path grows rapidly with energy
$\lambda_{\rm M.F.P.}\propto E^2$ \citep{revilleetal08,zirakashviliptuskin08}. Thus, the generation of 
large scale field structure is essential to achieve sub-Bohm diffusion at $E>10^{15}$~eV. The filamentation 
instability can grow extremely rapidly once the magnetic field perturbations have been driven to non-linear 
levels, and may help to significantly reduce the mean-free paths of these particles.  
Simulations in 3D, with a significantly larger dynamic range 
are required to confirm this.

The topic of cosmic rays escaping the source has been reinvestigated in recent years in light of 
developments in our understanding of magnetic field amplification 
\citep[e.g.][]{ohiraetal10,capriolietal10,drury11}. These models 
typically assume a one-dimensional, or spherically symmetric model of cosmic-ray diffusion. 
Filamentation is almost certainly important in this situation, since unlike the early one-dimensional
trapping models \cite[e.g][]{kulsrudzweibel75,kulsrud79}, the expansion of low density cavities and self-focussed filaments
can in fact assist in the escape. This most likely occurs only far upstream, since 
closer to the shock, the currents in the filaments are susceptible to beam-hose type instabilities. Again, a
three-dimensional analysis is required to investigate this further, and will be addressed in future
work.

\section*{Acknowledgements}

{ The research leading to these results has received funding from the European Research Council under the 
European Community's Seventh Framework Programme (FP7/2007-2013) / ERC grant agreement no. 247039. 
BR thanks A. Spitkovsky and L. Sironi for assistance in developing the kinetic part of the 
numerical code, and also S. O'Sullivan, J. Kirk, K. Schure, C. Ridgers and M. Tzoufras for invaluable discussion.}

\bsp

\label{lastpage}

\end{document}